\documentclass[intlimits,twoside,a4paper]{article}

\usepackage{amsmath,amssymb}
\usepackage{graphicx}

\usepackage[T2A]{fontenc}
\usepackage[cp1251]{inputenc}

\usepackage{cmpj3}

\issue{2018}{21}{4}{43704}
\doinumber{10.5488/CMP.21.43704}

\title[Electron energy spectrum and oscillator strengths]%
{Electron energy spectrum and oscillator strengths of quantum transitions in double quantum ring nanostructure driven by electric field}
\author[O.M.~Makhanets, V.I.~Gutsul, A.I.~Kuchak]{O.M.~Makhanets\footnote{E-mail: ktf@chnu.edu.ua}\,,
V.I.~Gutsul, A.I.~Kuchak}
\address{Yuriy Fedkovych Chernivtsi National University, 2 Kotsyubinsky St.,
58012 Chernivtsi, Ukraine}

\date{Received July 31, 2018, in final form September 26, 2018}

\begin{document}

\maketitle

\begin{abstract}
The effect of homogeneous electric field on the energy spectrum, wave functions of electron and oscillator strengths of intra-band quantum transitions in a double cylindrical quantum ring (GaAs/Al$_{x}$Ga$_{1-x}$As) is studied within the approximations of effective mass and rectangular potentials. The calculations are performed using the method of expansion of quasiparticle wave function over a complete set of cylindrical wave functions obtained as exact solutions of Schr\"odinger equation for an electron in a nanostructure without  electric field.
It is shown that the electric field essentially affects the electron localization in the rings of a nanostructure. Herein, the electron energies and oscillator strengths of intra-band quantum transitions non-monotonously depend on the intensity of electric field.

\keywords nanoring, electron, energy spectrum, oscillator strength, electric field

\pacs 73.21.La, 78.67.Hc
\end{abstract}

\section{Introduction}

Multilayered semiconductor nanostructures are studied both theoretically and experimentally for quite a long time. Unique properties of quasiparticles in such structures allow us to use them as the basic elements of modern nanoelectronic devices, such as tunnel diodes, lasers and detectors \cite{1,2,3}.

Semiconductor quantum rings occupy a separate place among  various types of nanostructures. As a rule, they have cylindrical symmetry as well as quantum wires \cite{4}. However, unlike the latter, their height is finite and can be of several nanometers. Therefore, the current of charge carriers in such nanostructures will be confined in all three dimensions and, in this respect, they are similar to quantum dots. Modern experimental possibilities allow one to grow nanoheterostructures with double cylindrical quantum rings on the basis of GaAs/Al$_{x}$Ga$_{1-x}$As semiconductors \cite{5,6,7}.

Cylindrical semiconductor quantum rings are intensively investigated theoretically \cite{8,9,10,11,12,13,14}. Changing the geometric sizes of the rings one can affect the energy spectra of quasiparticles and obtain the necessary optical properties. The external fields also essentially influence the spectra. In \cite{8,9},  the influence of a magnetic field on the energy spectrum of the electron and on the oscillator strengths of its intra-band quantum transitions in GaAs/Al$_{x}$Ga$_{1-x}$As rings was studied. It was shown that the electron energies and the oscillator strength of intra-band quantum transitions non-monotonously depend on the induction of the magnetic field. Besides, there was observed an anti-crossing of energy levels of the same symmetry over the magnetic quantum number (the Aaronov-Bohm effect) and brightly expressed maxima and minima  in the dependences of oscillator strengths on induction.

In the papers \cite{10,11,12}, the authors investigated the effect of a homogeneous electric field on the optical properties of  quantum nanorings using the model of a parabolic potential. The wave function of an electron in an electric field was written as an expansion over a complete set of its wave functions in an infinitely deep potential well with a further solution of the corresponding secular equation.

In the proposed paper, we are going to study the structure similar to that of \cite{10,11,12}. However, an optimal model of confined potential is to be used with an orthonormal basis of cylindrical wave functions obtained in the model of finite potential.  The oscillator strengths of intra-band quantum transitions as functions of  the electric field intensity in a double quantum ring  GaAs/Al$_{x}$Ga$_{1-x}$As nanostructure are analyzed.

\section{Theory of electron energy spectrum and oscillator strengths of intra-band quantum transitions in double quantum ring nanostructure \\driven by electric field}

\begin{figure}[!b]
\centerline{\includegraphics[width=0.5\textwidth]{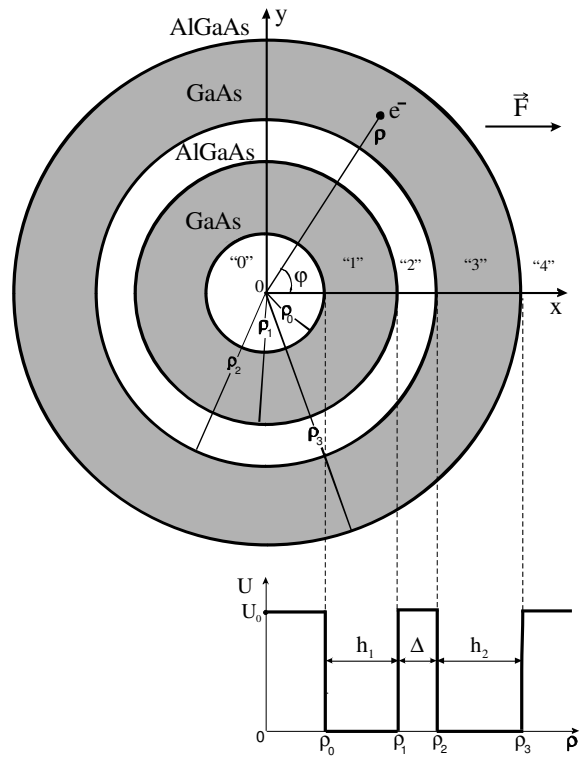}}
\caption{Cross-section of nanostructure, at height $L=5$~nm, by plane $z=0$ and potential energy profile.}
\label{fig1}
\end{figure}

The nanostructure consisting of two concentric rings (quantum wells GaAs) separated by a concentric and tunnel transparent ring Al$_{x}$Ga$_{1-x}$As of the width $\Delta $ is studied. The heights of three constructing parts are $L$. The inner and outer radii of the first ring are $\rho _{0} $ and $\rho _{1} $, its width is $h_{1} $, while those of the second ring are $\rho _{2} $ and $\rho _{3} $, respectively, its width is $h_{2} $. The cross-section by the plane $z=0$ and potential energy scheme of the structure is shown in figure~\ref{fig1}. A vector of electric field intensity $\vec{F}$ is directed along $Ox$ axis.

 According to the symmetry considerations, the further calculations are performed in a cylindrical coordinate system ($\rho $, $\varphi $, $z$) with $Oz$ axis directed along the axial axis of  the rings.
The electron effective masses are fixed in all parts of the structure
\begin{equation} \label{GrindEQ__1_}
\mu (\vec{r})=\left\{\begin{array}{l} {\mu _{0}\, ,\, \, \, |z|>L/2\, \ \  \text{or}\, \, \, |z|\leqslant  L/2\,  \ \  \text{and}\, \ \ \  0\leqslant  \rho \leqslant  \rho _{0}\, ,\, \, \, \, \rho _{1} \leqslant  \rho \leqslant  \rho _{2}\, , \ \ \ \, \, \rho >\rho _{3}\,, } \\
{\mu _{1}\, ,\, \, \, |z|\leqslant  L/2\,  \ \ \, \text{and}\,  \ \ \, \rho _{0} <\rho <\rho _{1}\, ,\, \, \, \rho _{2} <\rho \leqslant  \rho _{3}. } \end{array}\right. 
\end{equation}
In order to study the electron energy spectrum, there is solved the Schr\"odinger equation
\begin{equation} \label{GrindEQ__2_}
\hat{H}\Psi (\rho ,\varphi ,z)=E \Psi  (\rho ,\varphi ,z)
\end{equation}
with Hamiltonian
\begin{equation} \label{GrindEQ__3_}
\hat{H}=\frac{1}{2\mu (\vec{r})} \left[-\hbar ^{2} \left(\frac{\partial ^{2} }{\partial \rho ^{2} } +\frac{1}{\rho } \frac{\partial }{\partial \rho } +\frac{1}{\rho ^{2} } \frac{\partial ^{2} }{\partial \varphi ^{2} } +\frac{\partial ^{2} }{\partial z^{2} } \right)\right]-\frac{\hbar ^{2} }{2\mu (\vec{r})} \frac{\partial ^{2} }{\partial z^{2} } +U(\vec{r}) -|e| F\rho  \cos \varphi,
\end{equation}
where $e$ is the electron charge, $F$ is the magnitude of the electric field intensity, $U(\vec{r})$ is the potential of size quantization.

 Taking into account that the electric field does not influence the energy spectrum of the electron moving along $Oz$ axis and that the electron is mostly located in the quantum wells weakly penetrating into the barriers, the potential energy $U(\vec{r})$ is conveniently written as a sum
\begin{equation} \label{GrindEQ__4_}
U(\vec{r})=U(z)+U(\rho ),
\end{equation}
where
\begin{equation} \label{GrindEQ__5_}
U(z)=\left\{\begin{array}{l} {U_{0}\, ,\, \, \, |z|>L/2,} \\ {0,\, \, \, \, \, \, \, |z|\leqslant  L/2,} \end{array}\right. \qquad   U(\rho )=\left\{\begin{array}{l} {U_{0}\, ,\, \, \, 0\leqslant  \rho \leqslant  \rho _{0}\, ,\, \, \, \, \, \rho _{1} \leqslant  \rho \leqslant  \rho _{2} \,,\, \, \rho >\rho _{3}\, ,} \\ {0,\, \, \, \, \, \, \, \rho _{0} <\rho <\rho _{1} \,,\, \, \, \rho _{2} <\rho \leqslant  \rho _{3} .} \end{array}\right.
\end{equation}
In this case, $z$ variable is separated in a Schr\"odinger equation with Hamiltonian \eqref{GrindEQ__3_}, and the wave function can be written as follows:
\begin{equation} \label{GrindEQ__6_}
\Psi \left(\vec{r}\right)=\Phi (\rho ,\, \varphi )\, f(z).
\end{equation}

The Schr\"odinger equation for the electron moving along $Oz$ axis is easily solved \cite{15}. The wave functions $f(z)$ are obtained in the form
\begin{equation} \label{GrindEQ__7_}
\, f(z)=\left\{\begin{array}{l} {\left\{\begin{array}{l} {A^{(+)}  \cos (k_{0}  z)}, \\ {A^{(-)}  \sin (k_{0}  z)}, \end{array}\right. \, \, \, \, \, \, \, 0\leqslant  z\leqslant  L/2}, \\ {B \exp (-k_{1}  z), \, \, \, \, \, \, \, \ \ \  \, \, \, \, \, \, \, z>L/2.} \end{array}\right. 
\end{equation}
Using the condition of continuity of the wave function $f(z)$ and its density of current at the interface $z=L/2$ together with the normality condition, the unknown coefficients ($A^{(\pm )}, B$) are found and the dispersion equations too:
\begin{equation} \label{GrindEQ__8_}
\frac{k_{0} }{\mu _{0} } \tan\left(k_{0} \frac{L}{2} \right)=\frac{k_{1} }{\mu _{1} } \,,   \ \ \ \ \ \  \frac{k_{0} }{\mu _{0} } \cot\left(k_{0} \frac{L}{2} \right)=-\frac{k_{1} }{\mu _{1} }\, ,
\end{equation}
where $k_{0} =\sqrt{2\mu _{0} E_{n_{z} } /\hbar ^{2} } $, $k_{1} =\sqrt{2\mu _{1} (U_{0} - E_{n_{z} } )/\hbar ^{2} } $. The energy spectrum ($E_{n_{z} } $) of the electron moving along $Oz$ axis is obtained from equations \eqref{GrindEQ__8_} with quantum number $n_{z} $ numerating their solutions.

If there is no electric field ($F=0$), the Schr\"odinger equation with Hamiltonian \eqref{GrindEQ__3_} is solved exactly
\begin{equation} \label{GrindEQ__9_}
\Phi _{ n_{\rho }  m}^{ 0} (\rho , \varphi )=\frac{1}{\sqrt{2\piup } } R_{ n_{\rho } m} (\rho ) \re^{\ri m\varphi } .
\end{equation}
Here, $n_{\rho } $ and $m$ are the radial and magnetic quantum number, respectively, and the radial wave functions are written as follows:
\begin{equation} \label{GrindEQ__10_}
R_{n_{\rho } m}^{(i)} (\rho ) =A_{n_{\rho } m}^{(i)} j_{m}^{(i)} (\chi \rho )+B_{n_{\rho } m}^{(i)} n_{m}^{(i)} (\chi \rho ),\, \, \, \, \, i=0,1,2,3,4,
\end{equation}
\begin{equation} \label{GrindEQ__11_}
j_{m}^{(i)} (\chi \rho )=\left\{\begin{array}{l} {I_{m} (\chi _{0} \rho ),\, \, \, \, \, i=0,2,4,} \\ {J_{m} (\chi _{1} \rho ),\, \, \, \, \, i=1,3,} \end{array}\right. 
\end{equation}
\begin{equation} \label{GrindEQ__12_}
n_{m}^{(i)} (\chi \rho )=\left\{\begin{array}{l} {K_{m} (\chi _{0} \rho ),\, \, \, \, \, i=0,2,4,} \\ {N_{m} (\chi _{1} \rho ),\, \, \, \, \, i=1, 3,} \end{array}\right. 
\end{equation}
where $J_{m}$, $N_{m} $ are cylindrical Bessel functions of the first and second kind, $I_{m}$, $K_{m} $ are modified cylindrical Bessel functions of the first and second kind, $\chi _{0} =\sqrt{2\mu _{0} (U_{0} -E_{n_{\rho } m}^{0} )/\hbar ^{2} } $, $\chi _{1} =\sqrt{2\mu _{1} E_{n_{\rho } m}^{0} /\hbar ^{2} } $.

All unknown coefficients $A_{n_{\rho} m }^{(i)}$, $B_{n_{\rho} m }^{(i)} $ (hence, the wave functions) and electron energies $E_{n_{\rho } m}^{0} $ are obtained from the conditions of continuity of wave functions (\ref{GrindEQ__10_})--(\ref{GrindEQ__12_}) and their densities of currents at the interfaces of nanostructure and normality condition for the radial wave function. As far as the wave function should be finite at $\rho =0$ and $\rho \to \infty $, it means that the coefficients $B_{n_{\rho } m}^{(0)} =0$, $A_{n_{\rho } m}^{(4)} =0$.

If the structure is driven by an outer electric field, the Schr\"odinger equation with Hamiltonian \eqref{GrindEQ__3_} cannot be solved analytically. In order to find the electron spectrum at $F\ne 0$, the unknown wave functions are written as an expansion over a complete set of wave functions (\ref{GrindEQ__9_})
\begin{equation} \label{GrindEQ__13_}
\Phi _{n} (\rho , \varphi )=\frac{1}{\sqrt{2\piup } } \sum _{n_{\rho } }\sum _{m}c_{n_{\rho } m}^{n} R_{n_{\rho } m} (\rho )\re^{\ri m\varphi }   .
\end{equation}

Setting the expansion \eqref{GrindEQ__13_} into Schr\"odinger equation, the secular equation is obtained
\begin{equation} \label{GrindEQ__14_}
\big|H_{n_{\rho } m, n'_{\rho } m'} -E_{n} \delta _{n_{\rho } ,n'_{\rho } } \delta _{m,m'} \big|=0,
\end{equation}
where the matrix elements $H_{n_{\rho } m, n'_{\rho } m'} $ are of the form:
\begin{equation} \label{GrindEQ__15_}
H_{n_{\rho } m, n'_{\rho } m'} =E_{n_{\rho } m} \delta _{n_{\rho } ,n'_{\rho } } \delta _{m,m'} +\left(\delta _{m', m+1} +\delta _{m', m-1} \right)\frac{e F}{2} \int _{0}^{\infty }R_{n_{\rho } m} (\rho ) R_{n'_{\rho } m'} (\rho ) \rho ^{2}  \rd\rho  .
\end{equation}

We should note that, as it is clear from \eqref{GrindEQ__13_} and \eqref{GrindEQ__14_}, the new states of electron at its transversal movement are characterized by only one quantum number $n$.

Thus, the problem of the energy spectrum $E_{n} $ and wave functions $\Phi _{n} (\rho , \varphi )$ is reduced to the calculation of eigenvalues and eigenvectors of the obtained matrix. Hence, the complete wave functions $\Psi _{n n_{z} } (\vec{r})$ \eqref{GrindEQ__6_} of an electron and its energy $E_{n n_{z} } =E_{n} +E_{n_{z} } $ become known. They make it possible to evaluate the oscillator strengths of intra-band optical quantum transitions using the formula from \cite{16} 
\begin{equation} \label{GrindEQ__16_}
F_{n n_{z} }^{n' n'_{z} } \sim (E_{n' n'_{z} } -E_{n n_{z} } ) \big| M_{n n_{z} }^{n' n'_{z} } \big|^{2} ,
\end{equation}
where
\begin{equation} \label{GrindEQ__17_}
M_{n n_{z} }^{n' n'_{z} } =\big\langle n' n'_{z}  \big|  \sqrt{\mu (\rho )}  e\rho \cos\varphi \big| n n_{z}  \big\rangle
\end{equation}
is the dipole momentum of the transition.

\section{Analysis of the results}

The electron energies and its oscillator strengths of intra-band quantum transitions are studied as functions of the electric field intensity ($F$) for the double quantum ring GaAs/Al$_{0.4}$Ga$_{0.6}$As nanostructure with the physical parameters $\mu _{0} =0.063m_{0} $, $\mu _{1} =0.096m_{0} $, $U_{0} =297$~meV ($m_{0}$ is mass of pure electron in vacuum);   $a_\text{GaAs} =5.65$~{\AA} is GaAs lattice constant. All spectral parameters were calculated at a quantum number $n_{z} =1$, that is why it is omitted further.

\begin{figure}[!t]
\centerline{\includegraphics[width=0.96\textwidth]{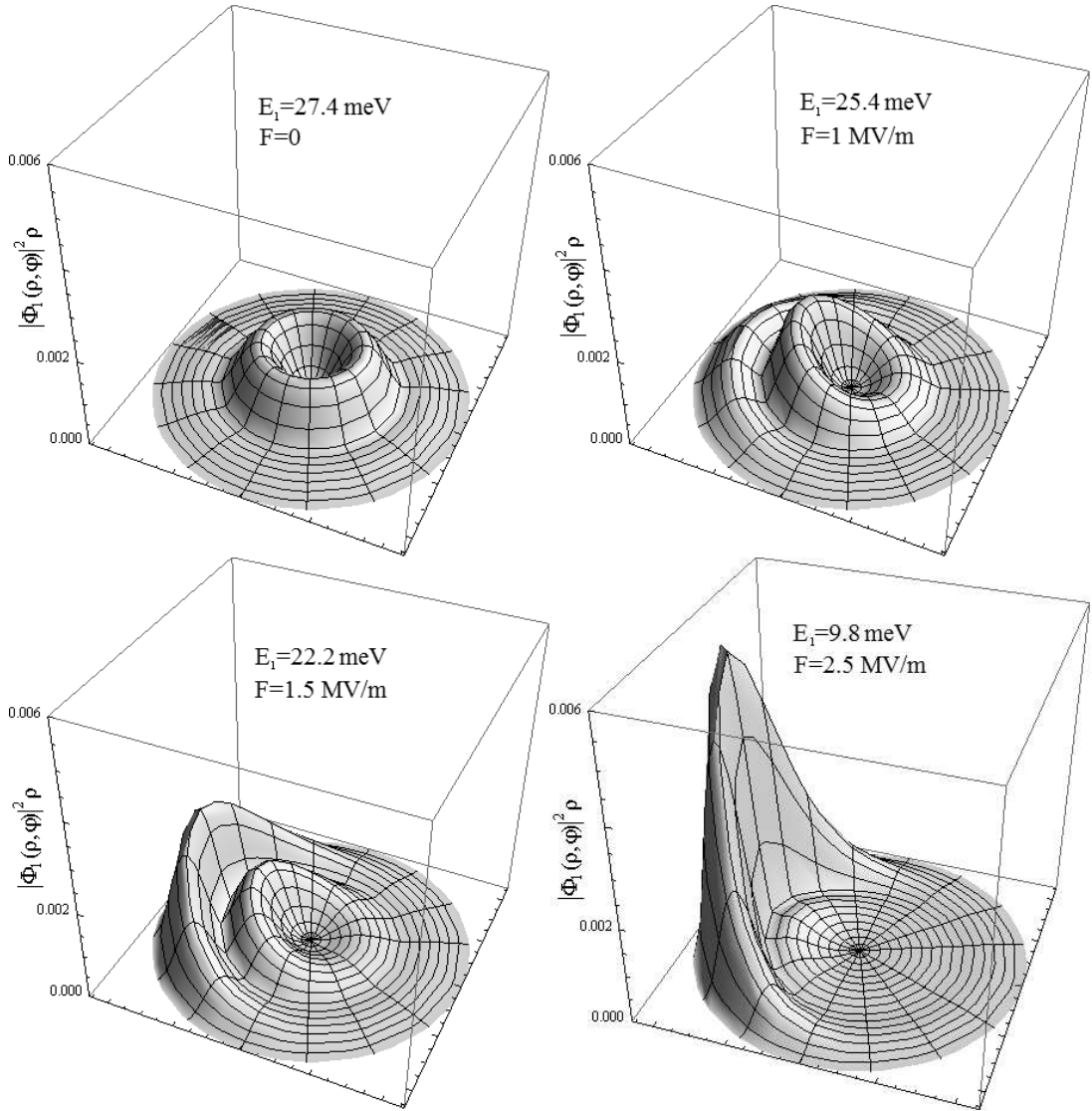}}
\caption{Probability density of electron (in ground state) location in a nanostructure  $\left|\Phi _{1} (\rho , \varphi )\right|^{2}  \rho $ at $L=5$~nm, $\rho _{0} =5 a_\text{GaAs} $, $h_{1} =18 a_\text{GaS} $, $\Delta =3 a_\text{GaS} $, $h_{2} =17 a_\text{GaS} $ and different electric field intensity: $F=0, 1, 1.5, 2.5$~MV/m.}
\label{fig2}
\end{figure}

In figure~\ref{fig2} the distribution of probability of electron (in ground state) location in nanostructure $\left|\Phi _{1} (\rho , \varphi )\right|^{2} \rho $ is shown at $L=5$~nm, $\rho _{0} =5 a_\text{GaAs} $, $h_{1} =18 a_\text{GaS} $, $\Delta =3 a_\text{GaS} $, $h_{2} =17 a_\text{GaS} $ and at different intensities of electric field: $F=0, 1, 1.5, 2.5$~MV/m. It is clear that an increasing intensity changes the location of electron in nanostructure. If $F=0$, it is located in the inner ring with the width $h_{1} $, while the angular distribution of probability is uniform. When intensity increases, the electron, in the ground state, tunnels from the inner ring into the outer ring in such a way that at $F=2.5$~MV/m it completely locates into the outer ring of the width $h_{2} $. Herein, its angular distribution essentially changes. The obtained result is in good qualitative agreement with the results of paper \cite{14}. If the  nanostructure studied is driven by a homogeneous magnetic field with the induction \textit{B} directed along $Oz$ axis, then, an increasing \textit{B} (at $F=0$) causes an increasing localization of the electron (in ground state) in the inner ring. Herein, the angular distribution of probability remains uniform \cite{9}.

\begin{figure}[!t]
\centerline{\includegraphics[width=0.88\textwidth]{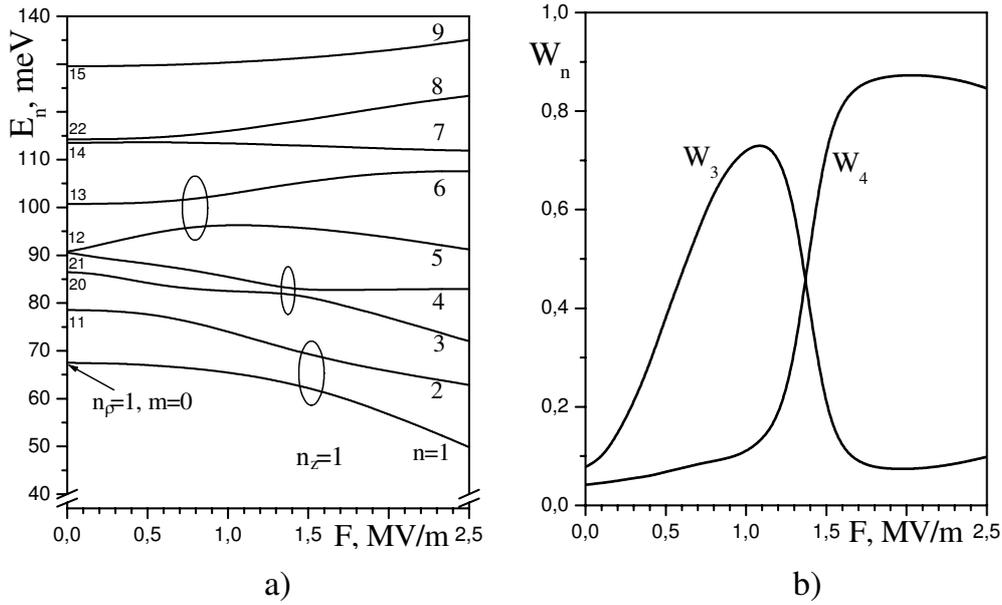}}
\caption{Electron energy $E_{n}$ (a) and complete probability $W_{n}$ of its location in the inner ring in quantum states $| 3\rangle $ and $| 4\rangle $ (b) as functions of the electric field intensity ($F$) at $L=5$~nm, $\rho _{0} =5 a_\text{GaAs} $, $h_{1} =18a_\text{GaS} $, $\Delta =3a_\text{GaS} $, $h_{2} =17a_\text{GaS} $.}
\label{fig3}
\end{figure}
\begin{figure}[!b]
\centerline{\includegraphics[width=0.65\textwidth]{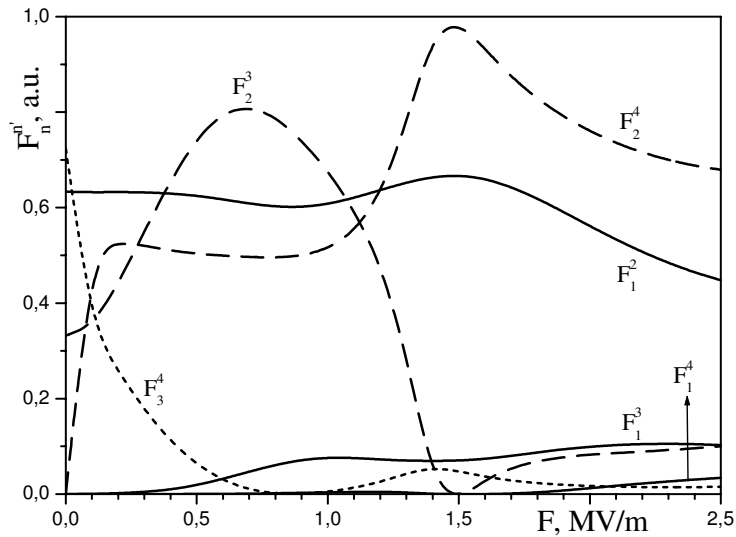}}
\caption{Oscillator strengths of intra-band quantum transitions as functions of the electric field intensity~($F$) at $L=5$~nm, $\rho _{0} =5 a_\text{GaAs} $, $h_{1} =18 a_\text{GaS} $, $\Delta =3 a_\text{GaS} $, $h_{2} =17 a_\text{GaS} $.}
\label{fig4}
\end{figure}

In figure~\ref{fig3}~(a) the electron energy ($E_{n} $) as a function of electric field intensity ($F$) is shown at $L=5$~nm, $\rho _{0} =5 a_\text{GaAs} $, $h_{1} =18 a_\text{GaS} $, $\Delta =3 a_\text{GaS} $, $h_{2} =17 a_\text{GaS} $. The figure proves that the ground state $({\left| 1 \right\rangle}) $ energy only decreases when $F$ increases. However, the energies of excited states demonstrate a different behaviour. In particular, the energy of the state ${\left| 5 \right\rangle} $ increases at first and then decreases. The energies of the states ${\left| 6 \right\rangle} $ and ${\left| 8 \right\rangle} $ increase in the whole interval of the intensity studied. Generally, an increase or a decrease of electron energies is determined by the location of the electron, in the corresponding state, in the particular ring and by the character of angular distribution of probability with respect to the direction of the electric field (for the ground state, figure~\ref{fig2}).

Since a potential barrier, which separates the rings, is of a finite height and width, then, the electron can tunnel from one quantum well into the other. This leads to a complicated and non-monotonous dependence of the electron energy spectrum on the intensity of the electric field. In particular, there are observed  anti-crossings of energy levels [for example, ${\left| 1 \right\rangle} $ and ${\left| 2 \right\rangle} $ at $F \sim1.5$~MV/m; ${\left| 3 \right\rangle} $ and ${\left| 4 \right\rangle} $ at $F \sim1.4$~MV/m; ${\left| 5 \right\rangle} $ and ${\left| 6 \right\rangle} $ at $F \sim0.8$~MV/m in figure~\ref{fig3}~(a)]. Anti-crossings appear depending on whether the electron, being in the neighbouring quantum states, is located in the outer or in the inner ring. This is well illustrated in figure~\ref{fig3}~(b), which shows the dependence of the complete probability ($W_{n} =\int\nolimits_{\rho _{0} }^{\rho _{1} } \int\nolimits_{0}^{2\piup }\left|\Phi _{n} (\rho , \varphi )\right|^{2} \rho  \rd\rho  \rd\varphi$) of electron location in the states ${\left| 3 \right\rangle} $ and ${\left| 4 \right\rangle} $ in the inner ring on the electric field intensity at the same geometrical parameters of the structure.

\begin{figure}[!b]
\centerline{\includegraphics[width=0.82\textwidth]{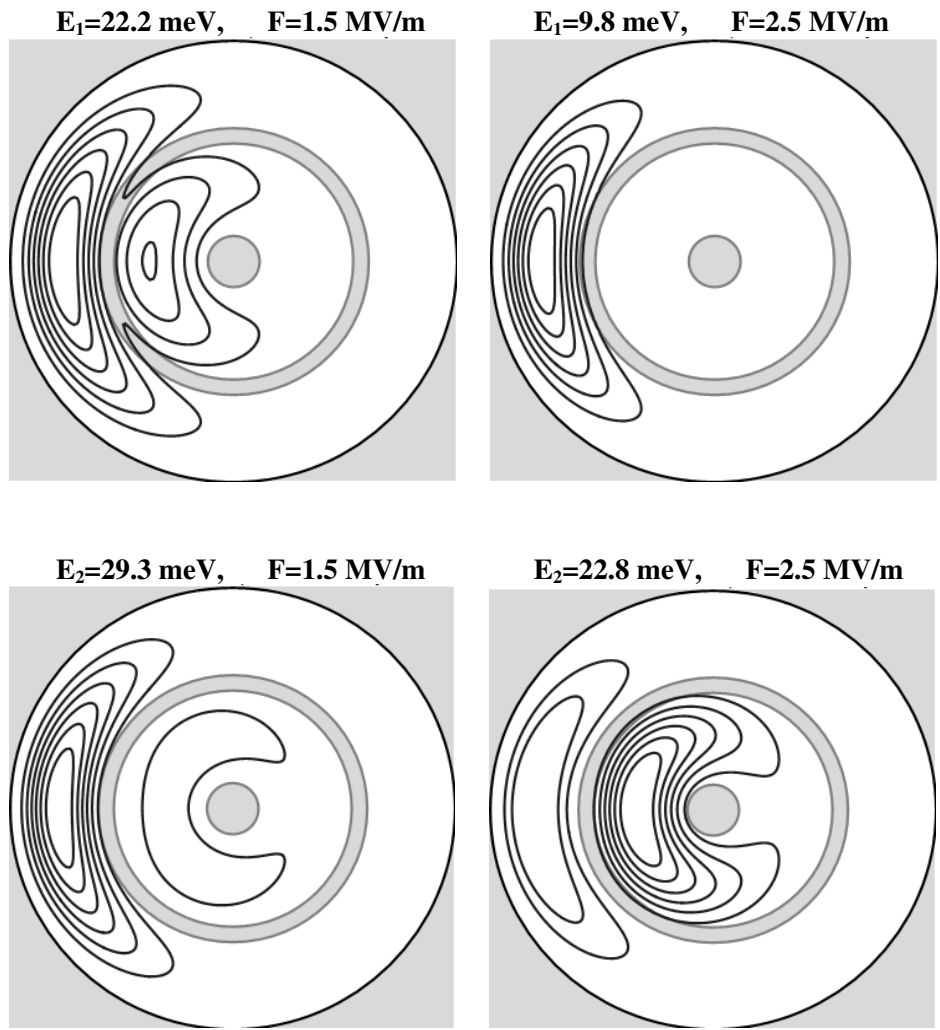}}
\caption{Contour of probability distribution of the electron location in a nanostructure in the states ${\left| 1 \right\rangle} $ and ${\left| 2 \right\rangle} $ at $F=1.5$~MV/m and $F=2.5$~MV/m, $L=5$~nm, $\rho _{0} =5 a_\text{GaAs} $, $h_{1} =18 a_\text{GaS} $, $\Delta =3 a_\text{GaS} $, $h_{2} =17 a_\text{GaS} $.}
\label{fig5}
\end{figure}

Figure proves that if $F=0$, the electron in both states ${\left| 3 \right\rangle} $ and ${\left| 4 \right\rangle} $ is located in the outer ring with a big probability. When the intensity increases, the probability of electron location in the inner ring increases for the state ${\left| 3 \right\rangle} $ and, at first, almost does not change for the state ${\left| 4 \right\rangle} $. At $F \sim1.1$~MV/m, $W_{3} $ approaches the maximum and rapidly decreases while $W_{4} $ rapidly increases. In the vicinity of $F \sim1.4$~MV/m, these probabilities become equal and an exchange of localization of the electron between the rings occurs for these states. Thus, the anti-crossing of energies $E_{3} $ and $E_{4} $ as functions of $F$ is observed [figure~\ref{fig3}~(a)]. We should note that such an effect is absent in a single nano-ring with one potential well. Moreover, it should be mentioned that quite similar series of levels, with respect to $n$ quantum number, will occur at $n_{z} =2, 3,\ldots,$ but they will be located in the high-energy region of the spectrum.

The capability of an electron, in different states, to be located in the inner ($h_{1} $) or the outer ($h_{2} $) ring causes a complicated and non-monotonous dependence of  oscillator strengths of intra-band quantum transitions on the intensity $F$ with brightly expressed maxima and minima (figure~\ref{fig4}). Herein, it turns out that such a non-monotonous behaviour of $F_{n }^{n'} $ is, mainly, determined by overlapping wave functions of the electron in the corresponding quantum states.

Let us observe, for example, the transition of an electron from the ground state ${\left| 1 \right\rangle} $ to the state~${\left| 2 \right\rangle} $ (curve $F_{1}^{2} $ in figure~\ref{fig4}) and the dependence of the probability density of the electron location in a nanostructure in these states ($\left|\Phi _{n} (\rho , \varphi )\right|^{2} \rho $) at $F=1.5$~MV/m, where the oscillator strength is maximal, and at $F=2.5$~MV/m, where it is minimal (figure~\ref{fig5}). Figure shows that at $F=1.5$~MV/m, the electron in the states ${\left| 1 \right\rangle} $ and ${\left| 2 \right\rangle} $ is mainly localized in the outer ring. Herein, the overlapping of the respective wave functions in formula \eqref{GrindEQ__17_} is essential and the oscillator strength is maximal though the difference of energies ($E_{2} -E_{1} $) is not big. When the electric field intensity increases, the difference of energies $E_{2} -E_{1} $ becomes bigger [figure~\ref{fig3}~(a)]. However, the electron in the state ${\left| 2 \right\rangle} $, under the influence of an electric field, begins to tunnel into the inner ring, which causes a smaller overlapping of the wave functions in \eqref{GrindEQ__17_} and, hence, a smaller oscillator strength of the respective transition. At $F=2.5$~MV/m, the electron in the state ${\left| 1 \right\rangle} $ is completely localized in the outer ring, while in the state ${\left| 2 \right\rangle} $ --- mainly in the inner ring (figure~\ref{fig5}). The wave functions of the corresponding states weakly overlap and the oscillator strength of the respective transition is small.

Quite similarly, due to the changes of location of the electron in the space of tunnel-connected quantum rings driven by an electric field, one can explain the non-monotonous behaviour of the oscillator strengths of quantum transitions between the other states.
 
Finally, we should note that an increasing nanostructure height \textit{L} causes a decrease of the electron energy $E_{n_{z} } $ at $n_{z} =1$. It tends to zero in the limit case ($\mathop{\lim }\nolimits_{L\to \infty } E_{n_{z} =1} =0$) and the spectrum $E_{n_{\rho }  m} $ completely corresponds to that in the structure consisting of two cylindrical nanotubes with an axial quasimomentum $k_{z} =0$.

\section{Conclusions}

\begin{enumerate}
\tolerance=3000%
\item  The electron energy spectrum and oscillator strengths of intra-band quantum transitions in a double quantum ring GaAs/Al$_{x}$Ga$_{1-x}$As nanostructure are studied as functions of the electric field intensity~($F$) within the approximations of effective mass and rectangular potentials.
\item  To calculate the energy spectrum and probability densities of the electron location in nano-rings driven by electric field, the stationary Schr\"odinger equation is solved using the method of expansion of a quasiparticle wave function over a complete set of wave functions in a nanostructure without the electric field.
\item  It is shown that the electric field essentially changes the distribution of probability of the electron location in a nanostructure. Thus, if the electron, in ground state, is located in the inner ring, then at an increasing electric field intensity, the quasiparticle tunnels into the outer ring.
\item  The electron energies and oscillator strengths of intra-band quantum transitions non-monotonously depend on the intensity of the electric field. One can observe anti-crossings of energy levels and brightly expressed minima and maxima in oscillator strengths as functions of $F$. Such a behaviour is caused by the change of the location of an electron, in different quantum states, in the space of two rings due to the varying electric field intensity.
\end{enumerate}

\ukrainianpart

\title{Енергетичний спектр електрона та сили осциляторів внутрішньозонних квантових переходів у подвійному нанокільці в електричному полі}
\author{О.М. Маханець, В.І. Гуцул, А.І. Кучак}
\address{Чернівецький національний університет імені Юрія Федьковича, \\вул. Коцюбинського, 2, 58012 Чернівці, Україна}

\makeukrtitle

\begin{abstract}
\tolerance=3000%
У моделі ефективних мас та прямокутних потенціалів досліджено вплив однорідного електричного поля на енергетичний спектр, хвильові функції електрона та сили осциляторів внутрішньозонних квантових переходів у подвійних напівпровідникових (GaAs/Al$_{x}$Ga$_{1-x}$As) циліндричних квантових кільцях. Розрахунки виконані методом розкладу хвильових функцій квазічастинки за повним набором циліндричних хвильових функцій, отриманих як точний розв'язок рівняння Шредінгера для електрона в наноструктурі за відсутності електричного поля.
Показано, що електричне поле суттєво впливає на локалізацію електрона у системі нанокілець. При цьому як енергії електрона,
так і сили осциляторів внутрішньозонних квантових переходів немонотонно залежать від величини напруженості електричного поля.
\keywords нанокільце, електрон, енергетичний спектр, сила осцилятора, електричне поле

\end{abstract}

\end{document}